\documentclass[aps,prb,twocolumn,groupedaddress,nofootinbib,notitlepage,showpacs,floatfix]{revtex4-1}

\usepackage{graphicx,graphics,epsfig,subfigure,times,bm,bbm,amssymb,amsmath,amsfonts,amsthm,mathrsfs,MnSymbol}
\usepackage[matrix,frame,arrow]{xypic}
\usepackage[pdfstartview=FitH]{hyperref}
\usepackage{pifont}
\hypersetup{
    colorlinks=true,       	% false: boxed links; true: colored links
    linkcolor=red,          	% color of internal links
   citecolor=magenta,        % color of links to bibliography
    filecolor=magenta,      	% color of file links
    urlcolor=cyan,           	% color of external links
    runcolor=cyan
}
\usepackage[pdftex]{color}
\usepackage{braket}%Dirac Notation in QM
\usepackage{enumerate}
\usepackage[normalem]{ulem}

\usepackage{subfigure}
\usepackage{overpic}

\usepackage{multirow}
\newcommand{\be}{\begin{equation}}
\newcommand{\ee}{\end{equation}}
\newcommand{\ba}{\begin{eqnarray}}
\newcommand{\ea}{\end{eqnarray}}
\newcommand\tr{{\mbox{Tr\,}}}
\newcommand{\ignore}[1]{}

\begin{document}

\title{Local characterization   of  1d topologically ordered states}

\author{Jian Cui$^{1,2}$, Luigi Amico$^{3,4}$, Heng Fan$^1$, Mile Gu$^{4,5}$, Alioscia Hamma$^{5,6}$, Vlatko  Vedral$^{4,7,8}$}
\affiliation{$^1$ Institute of Physics, Chinese Academy of Sciences, Beijing
100190, China}
\affiliation{$^2$ Freiburg Institute for Advanced Studies, Albert Ludwigs University of Freiburg, Albertstra{\ss}e 19, 79104 Freiburg, Germany}

\affiliation{ $^3$ CNR-MATIS-IMM $\&$
Dipartimento di Fisica e Astronomia,   Via S. Sofia 64, 95127 Catania, Italy}
\affiliation{ $^4$ Centre for Quantum Technologies, National University of
Singapore, 3 Science Drive 2, Singapore 117543}

\affiliation{$^5$Center for Quantum Information,
Institute for Interdisciplinary Information Sciences,
Tsinghua University, Beijing 100084, P.R. China}

\affiliation{$^6$Perimeter
Institute for Theoretical Physics, 31 Caroline St. N, N2L 2Y5,
Waterloo ON, Canada}

\affiliation{$^7$ Department of Physics, National University of
	Singapore, 2 Science Drive 3, Singapore 117542}
\affiliation{$^8$ Department of Physics, University of Oxford, Clarendon Laboratory, Oxford, OX1 3PU, UK}

\date{\today }

\begin{abstract}
We consider $1d$ Hamiltonian systems whose ground states {display symmetry protected topological order}.  We show that ground states within the topological phase cannot be connected {with} each other through LOCC  between a bipartition of the  system. Our claim is demonstrated by analyzing the entanglement spectrum and R\'{e}nyi entropies of different physical systems providing examples for symmetry protected topological phases. Specifically, we consider  spin-$1/2$ Cluster-Ising model and a class of spin$-1$ models undergoing quantum phase transitions to the Haldane phase.    Our results provide a  probe for symmetry-protected topological order. Since   the picture holds  even at the system's {\it local scale}, our analysis can serve as as  local experimental test for topological order.
\end{abstract}

\maketitle
\section{Introduction}
Understanding topological order in extended systems  is one of the major challenges in modern physics. Such an issue has immediate spinoff  in condensed  matter physics\cite{Fractional-Hall,topo_ins, White_liquids,Wen-1}, but encompasses important aspects of quantum information as well\cite{Sarma05}.  Topologically ordered systems are (generically) gapped systems characterized by a specific degeneracy  of the ground states\cite{Wen-1,Kitaev}.
The main attraction for  quantum computation applications relays on   the intrinsic robustness of such an order to external perturbations. Indeed such a property  is nothing but a  rephrasing that no local order parameter can be defined to  characterize topologically ordered states (Elitzur theorem).
This turns the main virtue of topological phases into a bottleneck, because any search for such kind of order in actual physical systems is problematic. {For the same reason, also more generic spin liquids, that may be not topologically ordered, but nevertheless possess a gap and no local order parameter, are long-sought states in nowadays condensed matter physics. \cite{White_liquids, Isakov}}

In the last few years,  topological order in many-body system has been studied resorting to new approaches. It is believed that  topological order can be characterized by the  entanglement encoded in the states of the system\cite{Amico_rev,Cramer_rev}. More precisely, it  is understood that  such an order is related to  a long range entanglement,  meaning that  topological states cannot be adiabatically connected   to non-topological ones resorting to  quantum circuits  made of  local (on the scale of the range of interaction in the underlying Hamiltonian) unitary gates\cite{gu-wen}.  Accordingly, any observable that would be able, in principle, to detect the topological order, is  intrinsically non local\cite{te,Renyi,Isakov}.

Here we push forward the idea that   progress can be made on this aspect, by analyzing how  topologically ordered ground states  change by varying the control parameter, within the quantum phase (instead of looking at a single copy of ground state corresponding to a fixed value of the control parameter)\cite{hamma_2dconvert}.
In particular, we  explore  whether the ground state `evolution' by changing the Hamiltonian's control parameter $p$ can be achieved by  LOCC, after having bi-partitioned system. The tool we exploit is provided  by the differential local convertibility\cite{local_convertibility}:
A given physical system is partitioned into two parties, $A$ and $B$,  limited to local operations on their subsystems  and classical inter-party
communication (LOCC). We comment that the notion of locality we refer does depend on the  partition that has been employed;  therefore the LOCC can indeed involve a portion of the system that can be very non-local on the scale fixed by the interactions in the  Hamiltonian. {This should be  contrasted with the protocols defining the topological order in terms of local unitary transformations mentioned above\cite{gu-wen}.} 

Assuming that $A$ and $B$ can share an entangled state (ancilla), the  differential local convertibility protocol can be feasibly expressed through a specific behavior of the  R\'{e}nyi entropies $\displaystyle{S_\alpha \doteq { {1}\over{1-\alpha} }  \log \tr \rho^\alpha }$. The differential convertibility holds if and only if $\partial_p S_\alpha \ge 0$, $\forall \alpha$, or $\partial_p S_\alpha \leq 0$, $\forall \alpha$\cite{local_convertibility}. In the present paper we will be using the latter  characterization. Such an approach was first  discussed in the realm of quantum critical phenomena  in the Ref.\cite{Cui_locc}.  In particular, we note that   paramagnets and phases with non vanishing local order parameters are indeed locally convertible\cite{Cui_comment}.

In this paper, we focus on spin systems in one spatial dimension, where topological order  is protected  by the symmetry of the system\cite{Wen-protected}.
We shall see that such symmetry-protected-topological-phases are not convertible. 
We shall see that such property holds on spatial scales that are smaller than the correlation length of the system.

To address the question,  we refer to specific spin systems providing paradigmatic examples in this context. First we will consider the cluster-Ising model\cite{Son11,smacchia} (see also\cite{Doherty09,Skrovseth09})). The physical platform for that is provided by cold atoms in a triangular optical lattice \cite{triangular,Pachos04}.   The model is particularly interesting in quantum information since it describes  how  the  $1d$-cluster states are quenched by  a qubit-qubit exchange interaction\cite{Briegel01}. Our second example is the $\lambda-D$ model, that is a  well known model for studying the Haldane order in $1d$ quantum magnets\cite{haldane,quantum_magnets}.
Quantum computation protocols based on Haldane-type states have been provided in\cite{akimasa}.

The main numerical tool for the analysis is  the DMRG technique with Matrix Product States  (MPS) variational ansatz\cite{mps-dmrg}. For both the cluster-Ising and the $\lambda$-D models, we analyze R\'{e}nyi entropies,  entanglement spectrum and differential local convertibility.
%As customarily, the DMRG results  we   will display are  reliable sufficiently far from the quantum phase transitions.

In sections \ref{convert_clusterising} and \ref{convert_lambda-D} the differential local convertibility of the cluster-Ising and $\lambda-D$ will be analysed. The scenario emerging from our study will be discussed in section \ref{discussion}. 
In the appendix \ref{app_clusterising},   \ref{app_lambda-D}, we discuss  edge states, correlation lengths and string order paramaters of the  models we deal with. The differential local convertibility for partition sizes larger than the correlation length of the system will be  discussed in appendix \ref{other_partitions}.
In the appendix \ref{large_alpha} we  discuss the R\'{e}nyi entropy in the  large $\alpha$ limit.

\section{The cluster-Ising model}
\label{convert_clusterising}
The Hamiltonian we consider is
\begin{equation}
H(g)=-\sum_{j=1}^N \sigma_{j-1}^x \sigma_j^z \sigma_{j+1}^x + g \sum_{j=1}^N \sigma_j^y \sigma_{j+1}^y,
\label{eq:ham_cluster}
\end{equation}
where $\sigma_i^{\alpha}$, $\alpha=x,y,z$, are the Pauli matrices and, except otherwise stated, we take open  boundary conditions $\sigma^{\alpha}_{N+1}= \sigma^{\alpha}_{0}=0$.
The phase diagram of (\ref{eq:ham_cluster}) has been investigated in \cite{Son11,smacchia}. For large  $g$ the system is an antiferromagnet with local order parameter. For $g=0$ the ground state is a a cluster state. It results  that the correlation pattern characterizing the cluster state  is robust up to a critical value of the control parameter, meaningfully defining a ``cluster phase" with  vanishing order parameter and string order\cite{Son11,smacchia}. Without symmetry, the cluster phase is a (non topological) quantum spin liquid, since there is a gap and no symmetry is spontaneously broken. Protected by a $Z_2 \times Z_2$ symmetry, the cluster phase is characterized by a topological fourfold ground state degeneracy, reflecting the existence of the edge states (see appendix \ref{app_clusterising})\cite{Son11,Son12}.  In the DMRG, we resolve the ground state degeneracy,  by adding a small perturbation $\sigma_1^x\sigma^z_2\pm\sigma^z_{N-1}\sigma_N^x$ to the Hamiltonian.

\begin{figure}
\includegraphics[width=0.85\columnwidth]{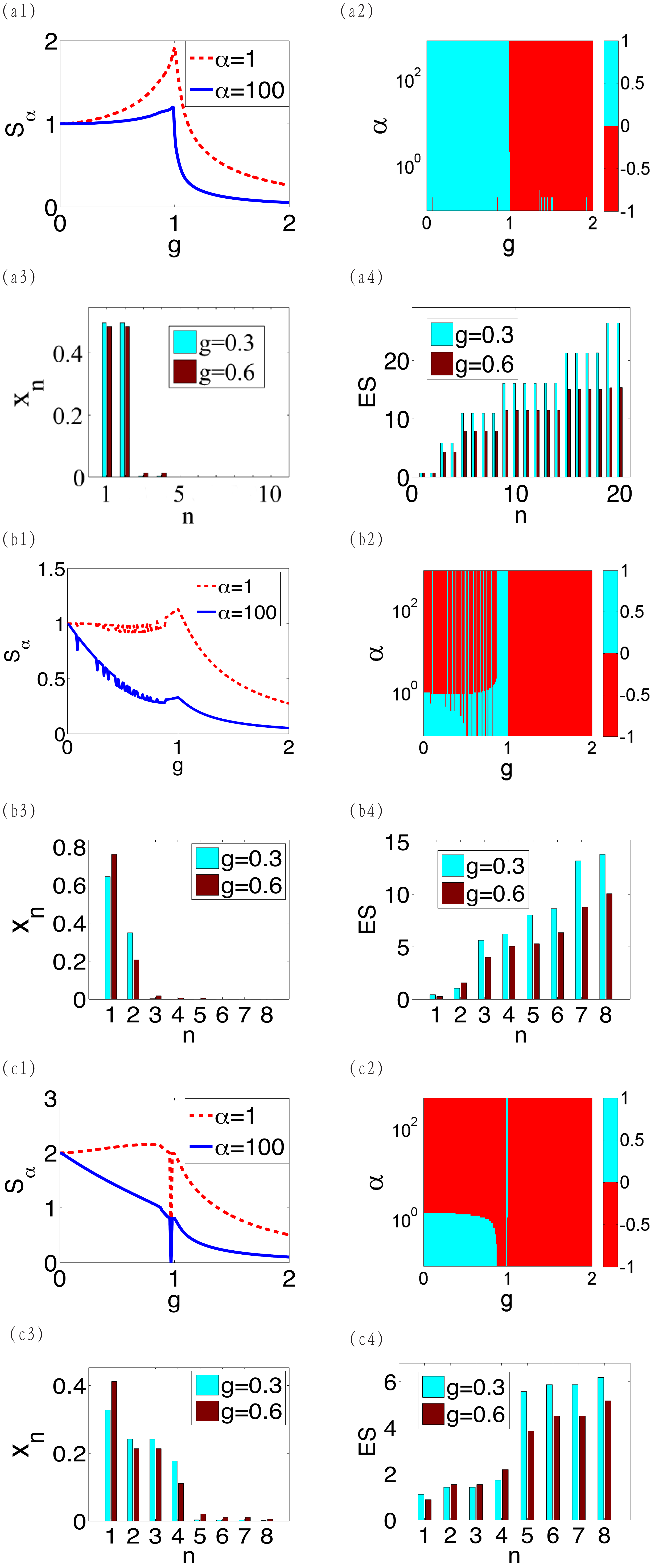}
\caption{The local convertibility  and the entanglement spectrum  of the cluster-Ising model Eq.(\ref{eq:ham_cluster}).
We characterize the differential local convertibility in terms of the the slopes of the R\'{e}nyi entropies.
Panel (a) is for bipartition $A|A$, $A=50$. There is differential local convertibility throughout the two different phases because,  for fixed $g$, $\partial_g S_\alpha$ does not change sign with $\alpha$.   Panel (b) is for bipartition $A|B$, $A=3$, $B=97$. Panel (c) is for $A|B|C$, being one blocks $A\cup C$ with $A=48$, $C=49$, and $B=3$.  In all of these cases, $\partial_g S_\alpha$ changes sign.
$(a3)\,,  (a4)$, $(b3)\,,  (b4)$, and $(c3)\,,  (c4)$ The larger and the smaller  eigenvalues  of reduced density matrix $x_n$, respectively;  $ES\doteq \left \{ -\log x_n\right \}$. In convertible phases, we observe that the change in the largest eigenvalues is 'faster' than the rate at which the smallest eigenvalues are populated. In contrast the  non differential local convertibility arises because the sharpening of the first part of the spectrum is over-compensated by the increasing of the smallest $x_n$.}
\label{convert_cluster}
\end{figure}

We find that the symmetric partition $A|A$ displays local convertibility, Fig.\ref{convert_cluster}: $(a1), (a2)$. This is indeed a fine tuned phenomenon since  the cluster phase results  non-locally convertible,  for a generic block of spins, both of the type $A|B$ and the  $B|A|B$,  Fig.\ref{convert_cluster}. We remark that such a property holds even for  size region $A$ smaller than the correlation length (see appendix  \ref{other_partitions} for other partitions).  Indeed, the entanglement spectrum is  doubly degenerate in all the cluster phase, as far as the size of the blocks $A$ and $B$ are larger than the correlation length. In contrast, the antiferromagnet is locally convertible, with non degenerate entanglement spectrum.
%\begin{figure}
%\includegraphics[width=\columnwidth]{figure1.pdf}
%\caption{The edge state, correlation length and the string order parameter of the cluster-Ising model.  $(a1)$ shows there is edge state in the cluster phase whereas there is no edge state in Ising antiferromagnetic phase. $(a2)$ shows  the correlation length of $\langle \sigma_n \sigma_{n+3}\rangle-\langle \sigma_n \rangle\langle\sigma_{n+3}\rangle$ displaying a critical behavior. $(a3)$ is the string order parameter ${\cal O}_z=(-)^{N-2}\langle \sigma_1^y\prod_{j=1}^{N-1} \sigma_j^z \sigma_N^y\rangle$.}
%\label{stat_cluster}
%\end{figure}

\section{The $\lambda-D$ model}
\label{convert_lambda-D}
In this section, we study the local convertibility of the   $\lambda-D$ model Hamiltonian describing {an interacting spin-$1$ chain} with  a single ion-anisotropy
\begin{eqnarray}
H=\sum_i[(S_i^xS_{i+1}^x+S_i^yS_{i+1}^y)+\lambda S_i^zS_{i+1}^z +D (S_i^z)^2].
\end{eqnarray}
where $S^u$, $u=\{x,y,z\}$ are spin-$1$ operators: $S^z|\pm\rangle=\pm |\pm \rangle$ and $S^z|0\rangle=0$.
The Hamiltonian above enjoys several  symmetries: the well known   $Z_2\times Z_2$,  and  the  link inversion  symmetry $S^u_j\rightarrow S^u_{-j+1}$ (see appendix \ref{app_lambda-D}).
The  phase diagram has been investigated by many authors\cite{lambdaDphasediagram,ercolessi,precise_dmrg} (see appendix \ref{app_lambda-D}) . Here we consider $\lambda > 0$. For small/large $D$ and fixed $\lambda$, the system is in a polarized state along $|+ \rangle \pm| - \rangle$ or $|0\rangle$, respectively. For large $\lambda$ and fixed $D$,  the  state displays  antifferomagnetic order.
At intermediate $D$ and $\lambda$, the state is a 'diluted anti-ferromagnet' with strong quantum fluctuations, defining the  Haldane phase, lacking of local order parameters and string order. For open boundary conditions (we apply in the present paper), the Haldane ground state displays a fourfold  degeneracy, that cannot be lifted without breaking the afore mentioned symmetry of the  Hamiltonian. This is the core mechanism defining the Haldane phase as a  symmetry-protected-topological ordered phase\cite{AKLT-pollmann,AKLT-protected}. {Without symmetry, the ground state is gapped and no symmetry is spontaneously broken, making the Haldane phase a quantum spin liquid.}
%
 %\begin{figure}
 %\label{sweeps}
%\includegraphics[height=4.5cm,width=\columnwidth]{figure2.pdf}
%\caption{We sweep through  the  phase diagram  in the two  following two ways: {\it 1)} fix $\lambda=1$ and change $D$; the Haldane phase is approximately located in the %range $-0.4\lesssim D\lesssim 0.8$.  {\it 2)} fix $D=0$, varying on $\lambda$; the Haldane phase is located in the range   $0\lesssim \lambda\lesssim 1.1$. }
%\label{sweeps}
%\end{figure}
%
%
%
%\begin{figure}
%\hspace*{-0.6cm}\includegraphics[width=1.1\columnwidth]{figure4.pdf}
%\caption{The  edge states, correlation lengths and string order parameters of the $\lambda-D$ model. The sweep (1) through the $\lambda-D$ phase diagram is considered (see text).
% In $(a1)$ we shows the Haldane phase edge states (see \cite{sorensen}); we do not find edge states in the other phases. In  $(a2)$ the string order parameters ${\cal O}_u=(-)^{N-2}\langle S_1^u\prod_{j=1}^{N-1} e^{i\pi S_j^u} S_N^{u}\rangle$. In   $(a3)$ the correlation length of $\langle  S_j^u S_{j+n}^u \rangle - \langle  S_j^u \rangle \langle S_{j+n}^u \rangle$. }
%\label{stat_lambdaD}
%\end{figure}
%
\begin{figure}
 \label{sweeps}
\includegraphics[height=5.5cm]{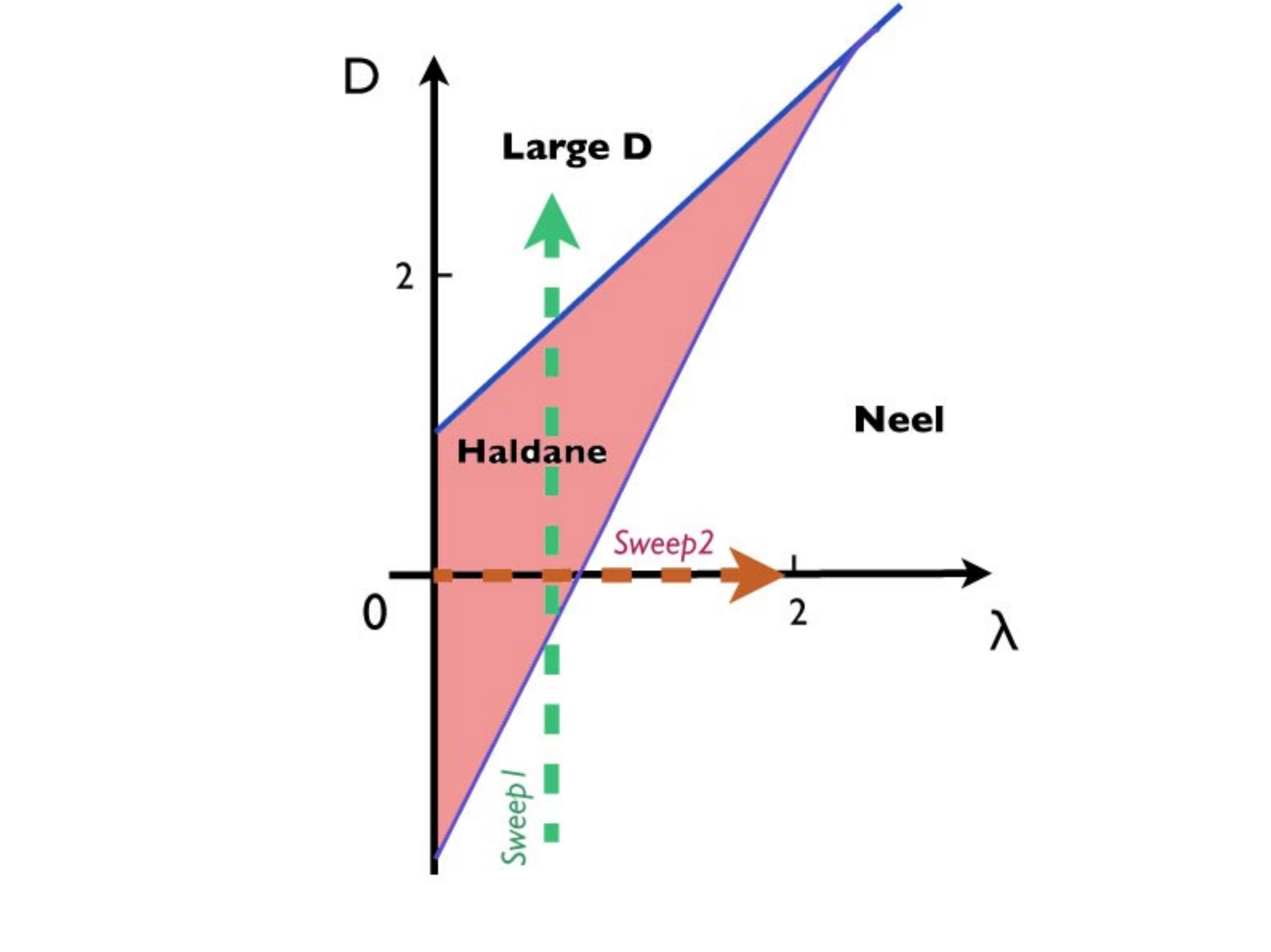}
\caption{We sweep through  the  phase diagram  in the two  following two ways: {\it 1)} fix $\lambda=1$ and change $D$; the Haldane phase is approximately located in the range $-0.4\lesssim D\lesssim 0.8$.  {\it 2)} fix $D=0$, varying on $\lambda$; the Haldane phase is located in the range   $0\lesssim \lambda\lesssim 1.1$. }
\label{sweeps}
\end{figure}
We sweep through  the  phase diagram  in the two  following two ways: {\it 1)} fix $\lambda=1$ and change $D$; the Haldane phase is approximately located in the range $-0.4\lesssim D\lesssim 0.8$.  {\it 2)} fix $D=0$, varying on $\lambda$; the Haldane phase is located in the range   $0\lesssim \lambda\lesssim 1.1$ (see Fig.\ref{sweeps}). 

 %We sweep through  the  phase diagram by: {\it 1)} fixing  $\lambda=1$ and change $D$;  {\it 2)} fixing  $D=0$ and  change  $\lambda$  (see Fig.\ref{sweeps}).
 We analyzed all the four states separately adding the  perturbation to the Hamiltonian
$\sim (S_1^z\pm S_N^z)$ with a small coupling constant.

\begin{figure}
\includegraphics[width=0.85\columnwidth]{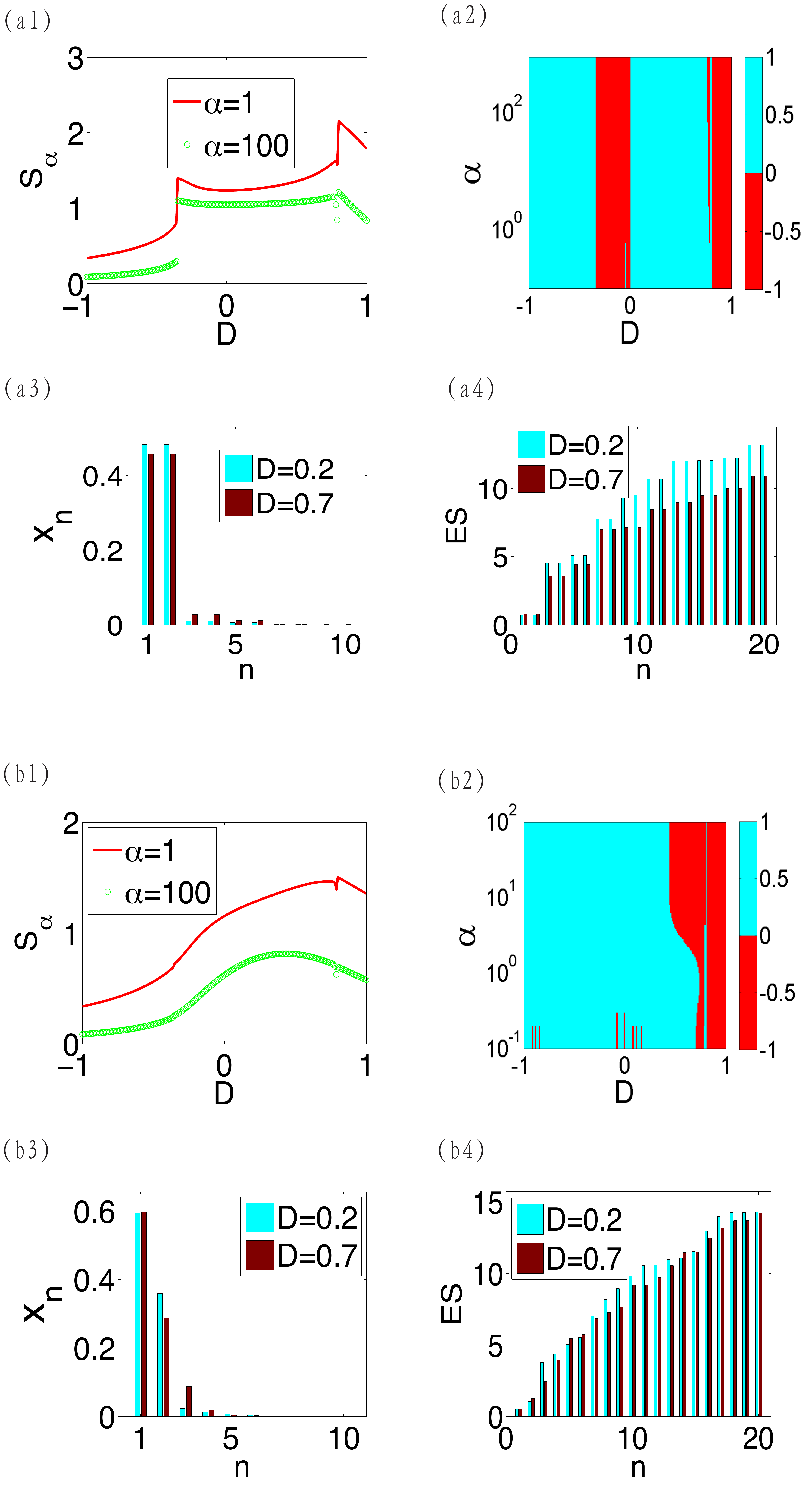}
\caption{The local convertibility for the partition $A|B$. The sweep (1) through the $\lambda-D$ phase diagram is considered (see also appendix \ref{app_lambda-D} for the schematic phase diagram).The upper panel displays the results for the symmetric case $A|A$.  The bottom panel refers to the antisymmetric case $A=96$, $B=4$. The Reny entropies  are presented in $(a1)$,  $(b1)$. The signs distributions of the derivatives of the R\'{e}nyi entropies  are shown in  $(a2)$, $(b2)$.  And eigenvalues of reduced density matrix $x_n$ and the entanglement spectrum are shown in $(a3)$,$(a4)$, $(b3)$, $(b4)$ as in Fig.\ref{convert_cluster}. The features of differential local convertibility are characterized by the slopes of the Reny entropies and correspond to specific features of the entanglement spectrum as explained in Fig.\ref{convert_cluster}.}
\label{conver_lambdaD_symm}
\end{figure}

We find that the N\'eel, ferromagnetic, and the large $D$ phases are locally convertible (seeFig.\ref{conver_lambdaD_symm}: $(a1)$,  $(a2)$).
Consistently with \cite{sym_protect_entang_spectr},
all of the Haldane ground states  are characterized by doubly degenerate entanglement spectrum for the symmetric $A|B$ partitions with $A=B$, for both sweep ways (Fig.\ref{conver_lambdaD_symm}:  $(a3)$ and $(a4)$) (see \cite{vondelft} for a recent progress on the understanding of double degenerate entanglement spectrum).  Such a property is not recovered both in the cases of asymmetric $A|B$ and $A|B|A$ partitions the entanglement spectrum is not found doubly degenerate, because the state of the system breaks the link inversion symmetry\cite{sym_protect_entang_spectr} (Fig.\ref{conver_lambdaD_symm}:  $(b3)$,  $(b4)$). See \cite{dechiara} for the analysis of the entanglement spectrum close to the quantum phase transitions.

We find that the Haldane phase is not locally convertible  (see  Fig.\ref{conver_lambdaD_symm}: $(b1)$, $(b2)$, \ref{fig_sweep1}, \ref{fig_sweep2}).
We remark that for both ways to partition the system the non-local-convertibility phenomenon is found even in the case of sizes of $B$ smaller than the correlation length  $\xi$(see \ref{app_lambda-D} for the behavior of $\xi$). {As for the model Eq.\ref{eq:ham_cluster},   we find that the symmetric bipartition $A=B$ displays local convertibility as a fine-tuned effect, that is broken for generic partitions (see appendix \ref{other_partitions} for other partitions).}
%
%
%%%
\begin{figure}[ht]
\includegraphics[width=\columnwidth]{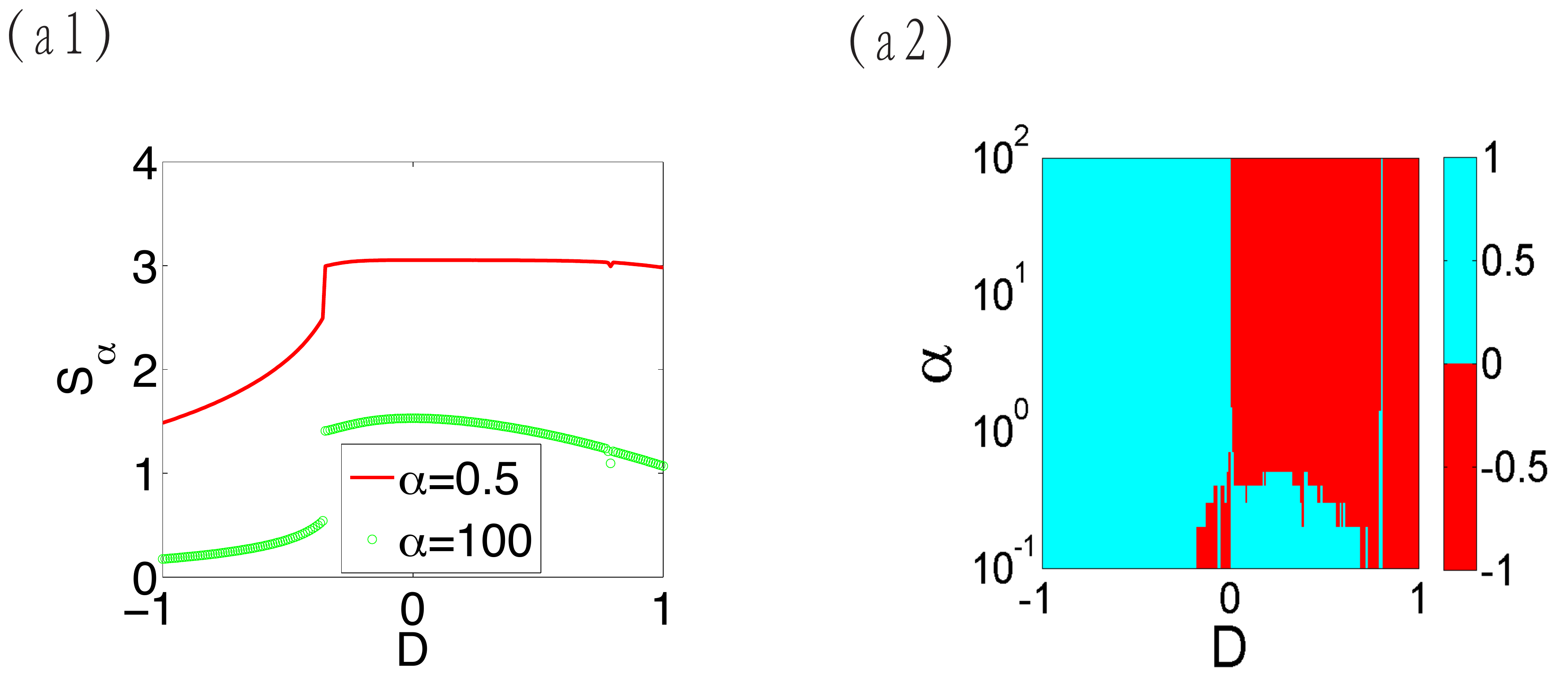}
\caption{\label{fig_sweep1} Sweep (1) through the $\lambda-D$ model: $\lambda=1$, $D\in \{-1,1\}$ (the schematic phase diagram is reported in appendix \ref{app_lamda-D}). The sign distribution of the derivative of  the R\'{e}nyi entropies $\partial_D S_\alpha$  for partitions $A|B|A$,  $A=48$ and $B=4$ ($N=100$) presented in $(a2)$. The features of differential local convertibility are characterized by the slopes of the Reny entropies and correspond to specific features of the entanglement spectrum as explained in Fig.\ref{convert_cluster}. The $S_\alpha$ are presented in $(a1)$ for  $\alpha=0.5,100$ decreasing from top to low.  All such quantitates are calculated for  the ground  state in $S_z^{tot}=1$ sector.}
\end{figure}
%
%%%
%
\begin{figure}[ht]
\includegraphics[width=\columnwidth]{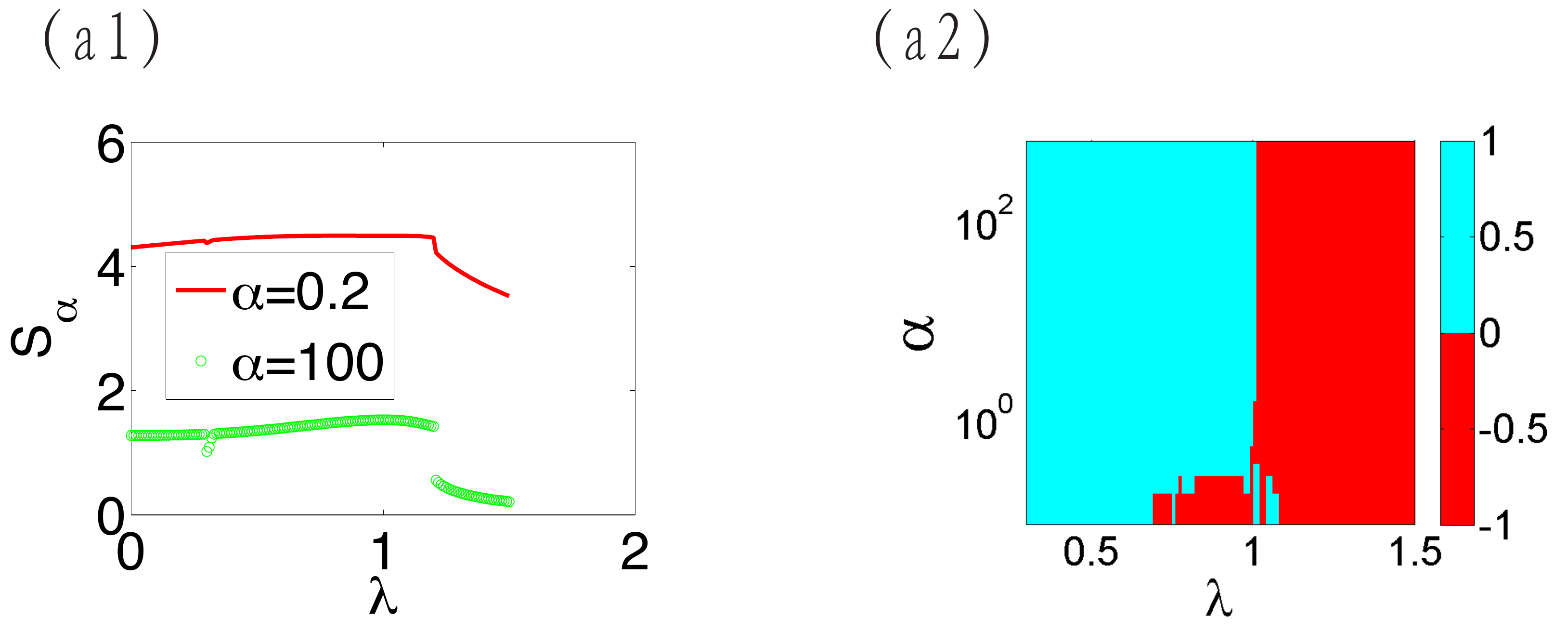}
\caption{\label{fig_sweep2} Sweep (2) through the $\lambda-D$ model: $D=0$, $\lambda\in \{0,1.5\}$ (see Fig.\ref{sweeps}  for a schematic phase diagram). The sign distribution of the derivativative of the R\'{e}nyi entropies $\partial_\lambda S_\alpha$  for partitions $A|B|A$,  $A=48$ and $B=4$ ($N=100$) presented in $(a2)$. The features of differential local convertibility are characterized by the slopes of the Reny entropies and correspond to specific features of the entanglement spectrum as explained in Fig.\ref{convert_cluster}. The $S_\alpha$ are presented in $(a1)$ for  $\alpha=100,0.2$ increasing from low to top.  All such quantitates are calculated for  the ground  state in $S_z^{tot}=1$ sector.
}
\end{figure}

\section{Discussion}
\label{discussion}
We explored quantitatively the notion of  LOCC in  $1d$-topologically  ordered systems.
In particular, we analyzed to what extent different ground states of the  Hamiltonian within the topological phase can be `connected' by (entanglement assisted)-LOCC between two parts $A$, $B$ in which the system has been divided. This issue is analyzed through the notion of differential local convertibility that can be expressed in terms of the properties of the Renyi entropies.  With this tool, we claim that progress can be made to detect quantum phases without local order parameter (contributing to the long-sought  hunt for  quantum spin liquids topologically ordered or not)\cite{hamma_2dconvert}.

In the present paper we specifically analyzed the  question  whether this method is useful to study topologically ordered phases assisted by  symmetries.
To this end,  we considered two very different models that are, at the same time, paradigmatic for the analysis of the notion of symmetry-protected-topological order:  the spin-$1/2$ cluster-Ising and the spin-$1$ $\lambda-D$ chains.
We find the symmetry-protected-topological phases are characterized by  non local convertibility, meaning that the Hamiltonian is more effective to drive the system through different topological states than LOCC between $A$ and $B$.
The phases with local order parameter turn out locally convertible.

The convertibility property is encoded in the specific response of  the entanglement spectrum  to the perturbation: The distribution of eigenvalues of the reduced density matrix get sharpened  in the most represented eigenvalues;  this implies  that high $\alpha-$R\'enyi entropies ca decrease;  at the same time, the spectrum acquires a tail made of small eigenvalues, because  more states are involved by increasing the  correlation length (see Fig.(\ref{convert_cluster}) $c2$, $c3$ and Fig.(\ref{conver_lambdaD_symm}) $b2$, $b3$)). The local convertibility is achieved, when the sharpening is compensated by the tail of the distribution
 (see Fig.\ref{convert_cluster}, Fig.\ref{conver_lambdaD_symm})\cite{note}.
 We remark the counter intuitive phenomenon for which some quantum correlations encoded in $S_\alpha$, for certain $\alpha$, decrease, in-spite of the increase in correlation length (see appendix \ref{app_lambda-D}).
Indeed,  similar   phenomenon was discovered in the $2d$ toric code \cite{hamma_2dconvert}, corroborating the scenario described in the present paper.  In particular, it was  noticed that, despite non topologically ordered,  also certain spin liquids result non-locally convertible for specific perturbations. Then, an interesting question that arises is concerning the 'stability' of the local convertibility by changing the perturbation. We observe that, on the other hand, that  spin liquids with some  symmetry protection,  are indeed stable, spanning  a well defined quantum phase, distinct from a paramagnet. The results of the present letter indicate that  the same symmetry protection is also able to protect their non local convertibility.

The  non local convertibility  occurs in the topological phases even for  subsystems sizes {\it smaller than the correlation length} of the system (see appendix \ref{app_lambda-D}, \ref{app_clusterising} for the behavior of $\xi$ in the models we analyzed); the degeneracy in the entanglement  spectrum, in contrast, is exhibited when the afore mentioned size is much larger than $\xi$.  Clearly, this paves the way to experimental tests through local measures on spatial region of sizes made of few spins, with the assistance of the protocols to address the R\'{e}nyi entropies provided recently\cite{exp_Reny}.
Incidentally, we note  that for the symmetric partition $A|B$ there is differential local convertibility, but this is a fine tuned effect that disappears if the two blocks have different size, see Fig.\ref{convert_cluster} and Fig.\ref{conver_lambdaD_symm}.  Ultimately, the size and type of subsystems $A$ and $B$ on which the differential local convertibility is displayed depends on the recombination of the edge states    that form at edges of the bipartition (see  \cite{Cui_comment} for an extensive discussion of such edge state recombination phenomenon).
With our results we can claim that the  local convertibility can characterize the phase, independently of  the way the system is partitioned.  We believe that such a scenario provides a valuable assistance to standard routes in experimental solid state physics to disclose topological order in the system.

Our  work opens several questions that will be subject of future investigation.
In particular, it is important to establish the precise relation between the ground state adiabatic evolution   and the differential local convertibility\cite{gu-wen}. Another interesting question is the role of differential convertibility in 2d symmetry protected topologically ordered systems like topological insulators.
%%%

\acknowledgments
We thank M.C. Banuls,  L. Cincio, F. Franchini, T. Shi, H.H. Tu, S. Santra and P. Zanardi for discussions.
This work was supported in part by the National Basic Research Program of China Grant 2011CBA00300, 2011CBA00301 the National Natural Science Foundation of China Grant 61073174, 61033001, 61061130540, 11175248, and the 973 program (2010CB922904).
Research at Perimeter Institute for Theoretical Physics
is supported in part by the Government of Canada through NSERC and
by the Province of Ontario through MRI.

\begin{appendix}

\section{ String order parameters, correlations length and edge states in Cluster-Ising}
\label{app_clusterising}

The cluster-Ising Hamiltonian  is
\begin{equation}
H(g)=-\sum_{j=1}^N \sigma_{j-1}^x \sigma_j^z \sigma_{j+1}^x + g \sum_{j=1}^N \sigma_j^y \sigma_{j+1}^y,
\label{eq:ham_cluster}
\end{equation}
Without symmetry, the cluster phase in Cluster-Ising model is a (non topological) quantum spin liquid, since there is a gap and no symmetry is spontaneously broken. Protected by a $Z_2 \times Z_2$ symmetry, the cluster phase is characterized by a topological fourfold ground state degeneracy in open boundary conditions, reflecting the existence of the edge states.
Such a degeneracy fans out from $g=0$ where $4$  Majorana fermions are left free at the free ends of the chain. The cluster phase can be characterized via a string order.
 The two phases are separated by a continuous quantum phase transition with central charge $c=3/2$.  Indeed the Hamiltonian (\ref{eq:ham_cluster}) is equivalent to three decoupled Ising chains \cite{Son11,smacchia}..

After the Jordan-Wigner transformation
$
\sigma_k^{+}=c_k^{\dag}\prod_{j<k}\sigma_j^z$, 
$\sigma_k^{-}=c_k\prod_{j<k}\sigma_j^z$, 
$\sigma_k^z=2c_k^{\dag}c_k-1$, 
the Hamiltonian of Cluster-Ising model can be written as
\begin{eqnarray}
H(g) 
%&=&-\sum_k(\sigma_k^x\sigma_{k+1}^z\sigma_{k+2}^x+g\sigma_k^y\sigma_{k+1}^y)\nonumber\\
%&=&-\sum_k\Big[(c_k^{\dag}+c_k)\prod_{j<k}\sigma_j^z\sigma_{k+1}^z\prod_{j<k+2}\sigma_j^z(c_{k+2}^{\dag}+c_{k+2})-g(c_k^{\dag}-c_k)\prod_{j<k}\sigma_j^z\prod_{j<k
%+1}\sigma_j^z(c_{k+1}^{\dag}-c_{k+1})\Big]\nonumber\\
%&=&-\sum_k\Big[(c_k^{\dag}+c_k)(2c_k^{\dag}c_k-1)(c_{k+2}^{\dag}+c_{k+2})-g(c_k^{\dag}-c_k)(2c_k^{\dag}c_k-1)(c_{k+1}^{
%\dag}-c_{k+1})\Big]\nonumber\\
%&=&-\sum_k\Big[(c_k-c_k^{\dag})(c_{k+2}^{\dag}+c_{k+2})+g(c_k^{\dag}+c_k)(c_{k+1}^{\dag}-c_{k+1})\Big]\nonumber\\
=-i \sum_k\Big[f_k^{(2)}f_{k+2}^{(1)}-gf_k^{(1)}f_{k+1}^{(2)}\Big].
\end{eqnarray}
where $f_k^{(1)}=c_k+c_k^{\dag}$ and  $f_k^{(2)}=-i(c_k-c_k^{\dag})$
are two different Majorana fermion operators.

\begin{figure}[h]
\includegraphics[width=\columnwidth]{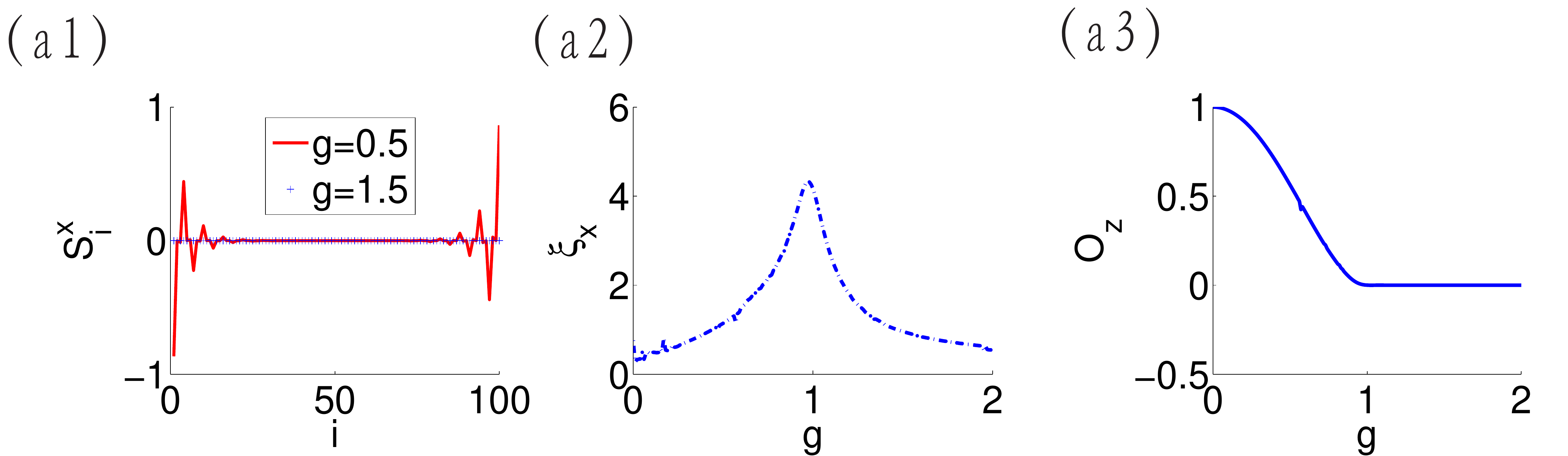}
\caption{The edge state, correlation length and the string order parameter of the cluster-Ising model.  $(a1)$ shows there is edge state in the cluster phase whereas there is no edge state in Ising antiferromagnetic phase. $(a2)$ shows  the correlation length of $\langle \sigma_n \sigma_{n+3}\rangle-\langle \sigma_n \rangle\langle\sigma_{n+3}\rangle$ displaying a critical behavior. $(a3)$ is the string order parameter ${\cal O}_z=(-)^{N-2}\langle \sigma_1^y\prod_{j=1}^{N-1} \sigma_j^z \sigma_N^y\rangle$.}
\label{stat_cluster}
\end{figure}

Although local order parameters do not exit to characterize the topological phase, the topological order in   Cluster-Ising model(see Fig.\ref{stat_cluster})  can be detected by the edge states (a1) and string order parameters (a3).

%\begin{figure}[h]
%\includegraphics[width=\columnwidth]{f2.pdf}
%\caption{Majorana fermion form of cluster-Ising
%chain with open boundary conditions. In the cluster limit where $g\rightarrow 0$, the first line and the 3rd line altogether leave 4 free Majorana fermions at the two %boundaries, which leads to the ground state degeneracy and edge state. Each separated line is exactly the Majorana fermion form of transverse filed Ising model.}
%\label{CI_Majorana}
%\end{figure}

\section{String order parameters, correlations length and edge states in  $\lambda-D$ model}
\label{app_lambda-D}
\begin{figure}[h]
\hspace*{-0.6cm}\includegraphics[width=1.1\columnwidth]{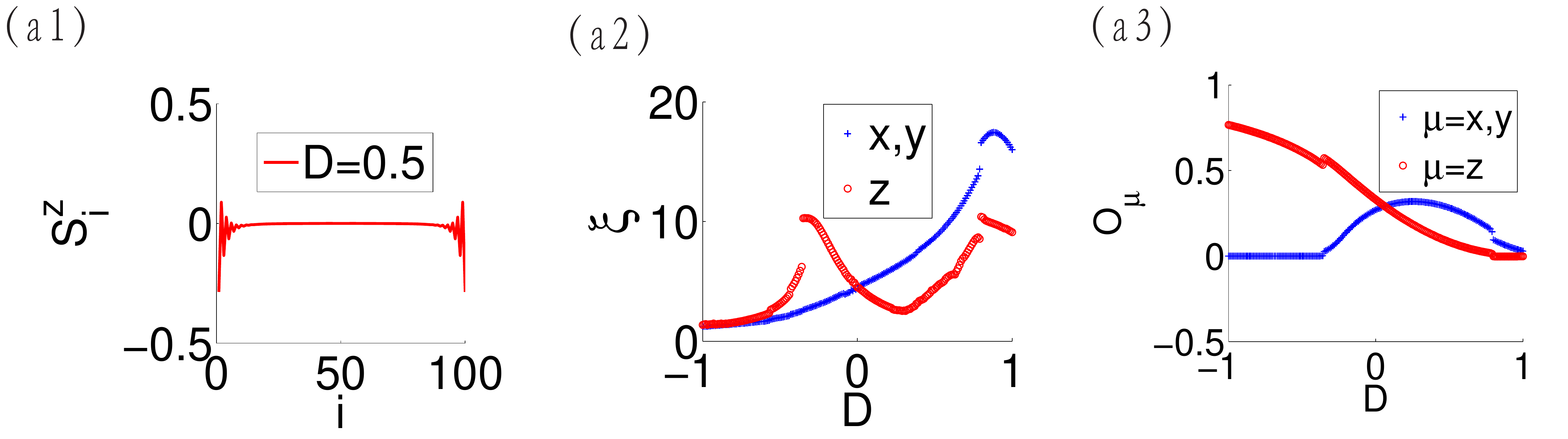}
\caption{The  edge states, correlation lengths and string order parameters of the $\lambda-D$ model. The sweep (1) through the $\lambda-D$ phase diagram is considered (see text).
 In $(a1)$ we shows the Haldane phase edge states; we do not find edge states in the other phases. In  $(a2)$ the string order parameters ${\cal O}_u=(-)^{N-2}\langle S_1^u\prod_{j=1}^{N-1} e^{i\pi S_j^u} S_N^{u}\rangle$. In   $(a3)$ the correlation length of $\langle  S_j^u S_{j+n}^u \rangle - \langle  S_j^u \rangle \langle S_{j+n}^u \rangle$. }
\label{stat_lambdaD}
\end{figure}

The  $\lambda-D$ Hamiltonian is 
\begin{eqnarray}
H=\sum_i[(S_i^xS_{i+1}^x+S_i^yS_{i+1}^y)+\lambda S_i^zS_{i+1}^z +D (S_i^z)^2].
\end{eqnarray}
The Hamiltonian above enjoys several  symmetries,  including  the  time reversal $S^{x,y,z} \rightarrow -S^{x,y,z} $, parity $S^{x,y} \rightarrow -S^{x,y} $,  $S^{z} \rightarrow S^{z} $ generating  $Z_2\times Z_2$,  and  the  link inversion  symmetry $S^u_j\rightarrow S^u_{-j+1}$.
For small/large $D$ and fixed $\lambda$, the system is in a polarized state along $|+ \rangle \pm| - \rangle$ or $|0\rangle$, respectively. For large $\lambda$ and fixed $D$,  the  state displays  antifferomagnetic order.
At intermediate $D$ and $\lambda$, the state is a 'diluted anti-ferromagnet' with strong quantum fluctuations, defining the  Haldane phase.
There is no local order parameters to characterize the Haldane phase in the $\lambda-D$ model, too. With symmetry protection, the topological order in Haldane phase can be detected by the edge states and string order parameters defined in Fig. \ref{stat_lambdaD} (see Ref.\ref{sorensen}). Without symmetry, the ground state is gapped and no symmetry is spontaneously broken, making the Haldane phase a quantum spin liquid. 
In Fig.\ref{sweeps}, we display the schematic phase diagram of the  $\lambda-D$.

\section{Differential local convertibility with subsystem size larger than or comparable with  the correlation length}
\label{other_partitions}

\begin{figure}
\includegraphics[width=\columnwidth]{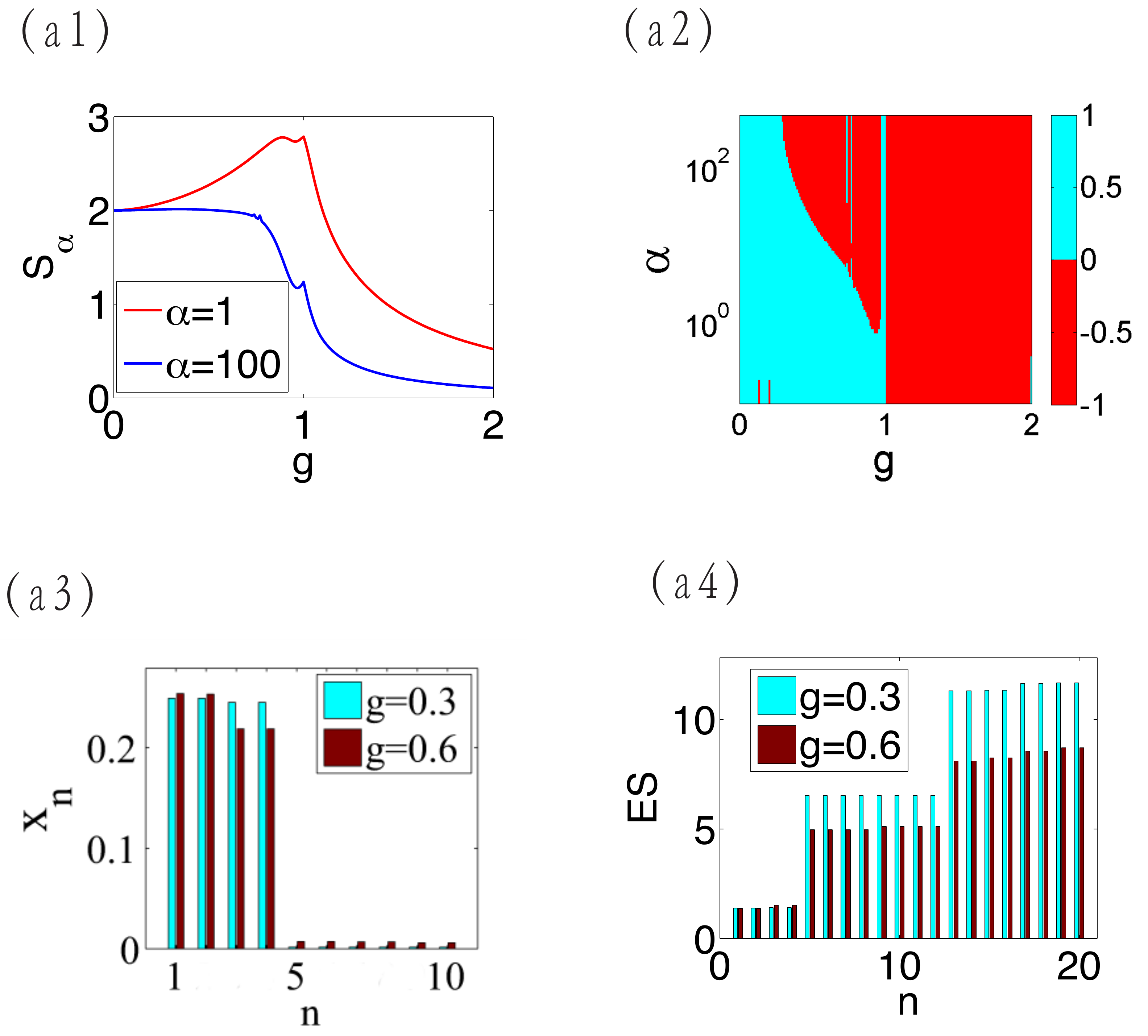}
\caption{The local convertibility  and the entanglement spectrum  of the cluster-Ising model with bipartition $45|10|45$. We characterize the differential local convertibility in terms of the the slopes of the R\'{e}nyi entropies.
  $\partial_g S_\alpha$ changes sign in the cluster phase.
$(a3)$  and $(a4)$ display  the largest and the smaller  eigenvalues  of reduced density matrix $x_n$, respectively;  $ES\doteq \left \{ -\log x_n\right \}$. In convertible phases, we observe that the change in the largest eigenvalues is 'faster' than the rate at which the smallest eigenvalues are populated. In contrast the  non differential local convertibility arises because the sharpening of the first part of the spectrum is over-compensated by the increasing of the smallest $x_n$.}
\label{convert_large_cluster}
\end{figure}

\begin{figure}[h]
\includegraphics[width=\columnwidth]{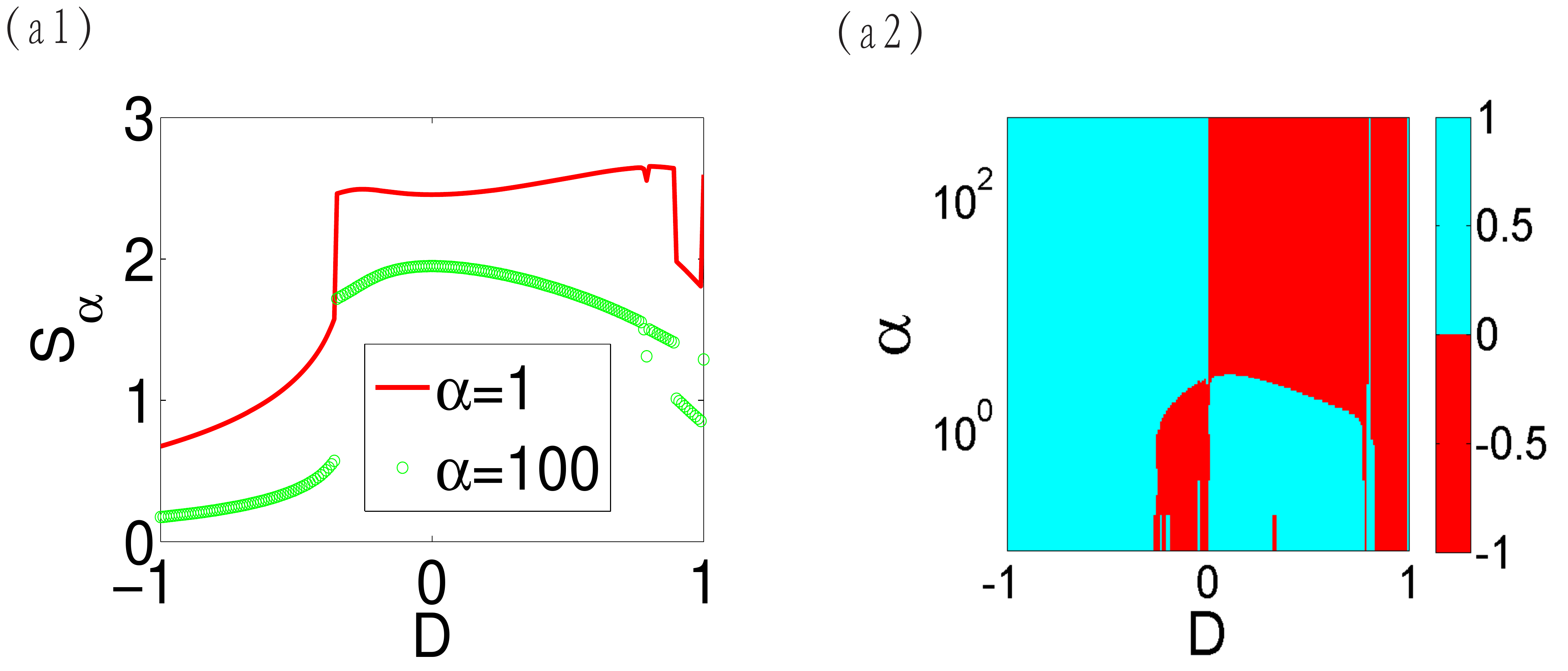}
\caption{\label{fig_sweep1} Sweep (1) through the $\lambda-D$ model: $\lambda=1$, $D\in \{-1,1\}$. The sign distribution of the derivative oconvertibilityf the R\'{e}nyi entropies $\partial_D S_\alpha$  for partitions $A|B|A$,  $A=45$ and $B=10$ ($N=100$) presented in $(a2)$. The features of differential local convertibility are characterized by the slopes of the Reny entropies and correspond to specific features of the entanglement spectrum as explained in Fig.\ref{convert_cluster}. The $S_\alpha$ are presented in $(a1)$ for  $\alpha=1,100$ decreasing from top to low.  All such quantitates are calculated for  the ground  state in $S_z^{tot}=1$ sector.}
\label{convert_large_lambda-D_1}

\end{figure}

\begin{figure}[h]
\includegraphics[width=\columnwidth]{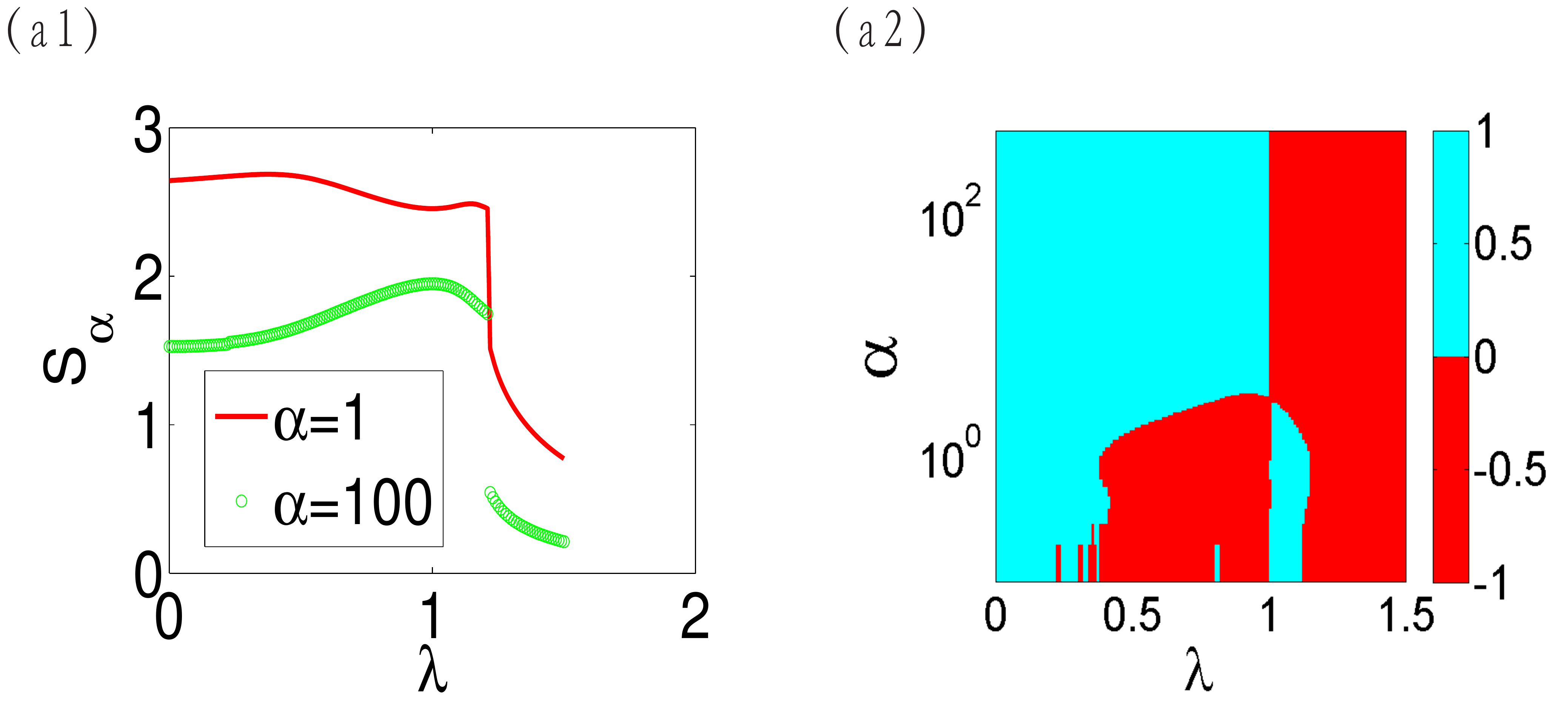}
\caption{\label{fig_sweep2} Sweep (2) through the $\lambda-D$ model: $D=0$, $\lambda\in \{0,1.5\}$ . The sign distribution of the derivativative of the R\'{e}nyi entropies $\partial_\lambda S_\alpha$  for partitions $A|B|A$,  $A=45$ and $B=10$ ($N=100$) presented in $(a2)$. The features of differential local convertibility are characterized by the slopes of the Reny entropies and correspond to specific features of the entanglement spectrum as explained in Fig.\ref{convert_cluster}. The $S_\alpha$ are presented in $(a1)$ for  $\alpha=100,1$ increasing from low to top.  All such quantitates are calculated for  the ground  state in $S_z^{tot}=1$ sector.
}
\label{convert_large_lambda-D_2}
\end{figure}

In the main text we have shown that differential local convertibility method works well with subsystem size smaller than the correlation length. In this section we present  the results of subsystem size larger or equivalent to the correlation length (Figs.\ref{convert_large_cluster}, \ref{convert_large_lambda-D_1}, \ref{convert_large_lambda-D_2}). 

%%%
%
%%%
%

\section{Large $\alpha$ limit of R\'{e}nyi entropy and local convertibility }
\label{large_alpha}
\begin{figure}[h]
\hspace*{-0.6cm}\includegraphics[width=1.1\columnwidth]{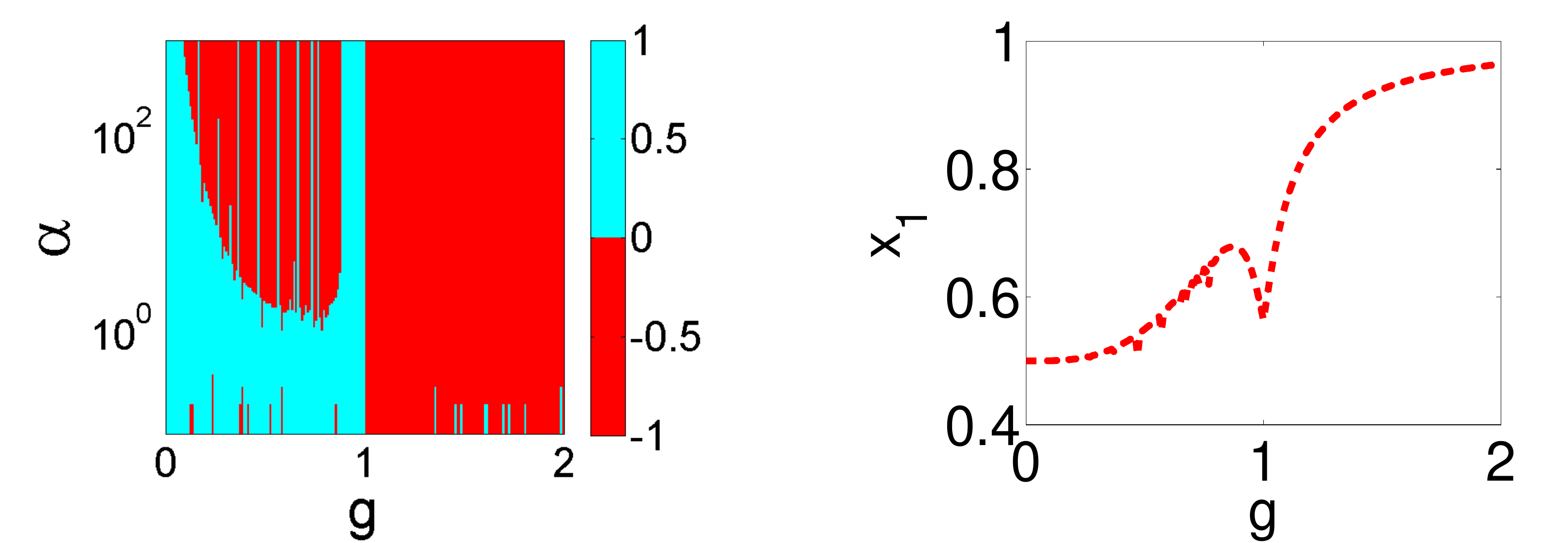}
\caption{Cluster-Ising model, $90|10$ bipartition. The left panel is the sign distribution of R\'{e}nyi entropy derivative which characterizes the region with non local convertibility. The right panel is the largest eigenvalue, whose slope has the opposite sign with the R\'{e}nyi entropy derivative in the large $\alpha$ limit. Comparing the two panels we can see that our conclusion is still correct even we go to infinite $\alpha$.  }
\label{lam_max}
\end{figure}

In the main text, by calculating the R\'{e}nyi entropies with different parameters, we have generally shown that with fixed bipartition of the spin chain, states with symmetry protected topological order cannot convert to each other via LOCC (assisted by entanglement), which is different from the stats with local order. Indeed, to arrive at such a conclusion rigorously, we have to calculate infinite R\'{e}nyi entropies with $\alpha$ from $0$ to $\infty$. From the definition of R\'{e}nyi entropy $\displaystyle{S_\alpha \doteq { {1}\over{1-\alpha} }  \log \tr \rho^\alpha }={ {1}\over{1-\alpha} }  \log \sum_i x_i^\alpha$ we can see that if we directly calculate the R\'{e}nyi entropy with very large $\alpha$ numerically, the numerator and denominator are both infinitely large such that computer cannot give correct results.
Therefore in the $\alpha \rightarrow \infty$ limit, we can apply the L'Hospital's rule to obtain $S_{\infty}=-\log x_1$, where $x_1$ is the largest eigenvalue. Notice that the $S_{\alpha}$ is a smooth and monotonic function of $\alpha$, therefore we can arrive at the rigorous conclusion numerically by going to the numeric limit of $\alpha$  assisted with the verification by $x_1$, see Fig. \ref{lam_max}

\end{appendix}


\begin{thebibliography}{99}
%
\bibitem{Fractional-Hall}
H. L. Stormer, D. C. Tsui, and A. C. Gossard, Rev. Mod. Phys. {\bf 71}, S298 (1999).
\bibitem{topo_ins} M. Z. Hasan and C. L. Kane, Rev. Mod. Phys. {\bf 82}, 3045  (2010).
\bibitem{White_liquids} S. Yan,  D. A. Huse, and  S. R.  White, Science, {\bf  332}, 1173 (2011).
\bibitem{Wen-1} X.-G. Wen, \emph{Quantum Field Theory of Many-body Systems}
(Oxford University Press, 2004).
\bibitem{Sarma05} M. H. Freedman, A. Kitaev, and Z. Wang, Commun. Math. Phys.
\textbf{227}, 587 (2002);  C. Nayak, S. H. Simon, Ady Stern, M. Freedman, S. Das Sarma, Rev. Mod. Phys. {\bf 80}, 1083 (2008).
\bibitem{Kitaev} A. Y. Kitaev, Ann. Phys. (N. Y.) \textbf{303}, 2 (2003).
\bibitem{Amico_rev} L. Amico,  R. Fazio, A. Osterloh, and V. Vedral, Rev. Mod. Phys. {\bf 80}, 517 (2008).
\bibitem{Cramer_rev} J. Eisert, M. Cramer, and M. B. Plenio, Rev. Mod. Phys. {\bf 82}, 277 (2010).
\bibitem{gu-wen}  Z.-G. Gu and X.G. Wen,  Phys. Rev. B {\bf 80}, 155131 (2009).
\bibitem{te} A. Hamma, R. Ionicioiu, and P. Zanardi, Phys. Lett. A
\textbf{337}, 22 (2005);  A. Hamma, R. Ionicioiu, and P. Zanardi,
Phys. Rev. A \textbf{71}, 022315 (2005);
 A. Kitaev and J. Preskill, Phys. Rev. Lett. \textbf{96},
110404 (2006); M. Levin and X.-G. Wen, Phys. Rev. Lett.  \textbf{96},
110405 (2006).
\bibitem{Renyi} S. T. Flammia, A. Hamma, T. L. Hughes, and X.-G.
Wen, Phys. Rev. Lett. \textbf{103}, 261601 (2009).
\bibitem{Isakov} S. V. Isakov, M. B. Hastings, and R. G. Melko, Nature Phys.
\textbf{7}, 772 (2011).
\bibitem{hamma_2dconvert} A. Hamma, L. Cincio, S. Santra, P. Zanardi, and L. Amico,  Phys. Rev. Lett. {\bf 110}, 210602 (2013).
\bibitem{local_convertibility} S. Turgut, J. Phys. A: Math. Theor. {\bf 40}, 12185 (2007); M. Klimesh,  arXiv:0709.3680 (2007).
\bibitem{Cui_locc} J. Cui, M. Gu, L.-C. Kwek, M. F. Santos, H. Fan, V. Vedral, Nat Commun {\bf 3}, 812  (2012).
\bibitem{Cui_comment}  We remark that both the paramagnet and the ordered phases turn out locally convertible if symmetry breaking is taken into account F. Franchini 
J. Cui, L. Amico, H. Fan, M. Gu, A. Hamma, V.E. Korepin, L.-C. Kwek, V. Vedral,  arXiv:1306.6685.
\bibitem{Wen-protected} X. Chen, Z-C. Gu, X-G. Wen, Phys. Rev. B {\bf 82}, 155138  (2010);  {\it ibid} Phys. Rev. B  {\bf 83}, 035107 (2011);  {\it ibid} arXiv:1103.3323  (2011).
\bibitem{Son11}W. Son, L. Amico, R. Fazio, A. Hamma, S. Pascazio and V. Vedral, Europhys. Lett., {\bf 95}, 50001  (2011).
\bibitem{smacchia}  P. Smacchia, L. Amico, P. Facchi, R. Fazio, G. Florio, S. Pascazio, and V. Vedral Phys. Rev. A  {\bf 84}, 022304 (2011).
\bibitem{haldane} F. D. M. Haldane, Phys. Rev. Lett. {\bf 61}, 1029-1032 (1988).
\bibitem{quantum_magnets}H.J. Mikeska and A.K. Kolezhuk,  {\it Quantum Magnetism}, U. Schollw\"ock, J. Richter, D.J.J. Farnell, R.F. Bishop  Eds (Springer, Berlin, 2004).
%
\bibitem{triangular}
	C. Becker, P. Soltan-Panahi, J. Kronj\"ager , S. D\"orscher, K. 	Bongs, and K. Sengstock,  New J. Phys. {\bf 12}, 065025 (2010).
\bibitem{Pachos04}
	J. K. Pachos and M. B. Plenio, {\prl} {\bf 93}, 056402 (2004).
\bibitem{Briegel01}
	H. J. Briegel and R. Raussendorf, {\prl} {\bf 86}, 910 (2001).
	\bibitem{akimasa} J. M. Renes, A. Miyake, G. K. Brennen, and S. D. Bartlett,  New J. Phys. {\bf 15}, 025020 (2013).
	\bibitem{mps-dmrg}F. Verstraete,  V. Murg, and J. I. Cirac, Adv. Phys.  {\bf 2}, 143 (2008).
\bibitem{Doherty09}
	A. C. Doherty and S. D. Bartlett, {\prl} {\bf 103}, 020506 (2009).
\bibitem{Skrovseth09}
	S. O. Skr{\o}vseth and S. D. Bartlett, {\pra} {\bf 80}, 022316 (2009).
\bibitem{lambdaDphasediagram}W. Chen, K. Hida, and B. C. Sanctuary, Phys. Rev. B. \textbf{67}, 104401 (2003).
\bibitem{ercolessi} C. Degli Esposti Boschi, E. Ercolessi, G. Morandi,  in {\it Symmetries in Science XI}, 145-173, (Kluwer 2004); arXiv:cond-mat/0309658.
\bibitem{precise_dmrg}  S. Hu,  B. Normand,  X.  Wang, and L Yu, Phys. Rev. B {\bf 84}, 220402(R) (2011).
\bibitem{Son12} W. Son, L. Amico, and V. Vedral, Quant. Inf. Proc. {\bf 11},  1961 (2012).
\bibitem{AKLT-pollmann}F. Pollmann, E. Berg, A. M. Turner, M. Oshikawa, Phys. Rev. B {\bf 85}, 075125 (2012).
\bibitem{AKLT-protected} S-P. Kou and X-G. Wen, Phys. Rev. B {\bf 80}, 224406 (2009).
\bibitem{sym_protect_entang_spectr}  F. Pollmann, E. Berg, A. M. Turner, M. Oshikawa  Phys. Rev. B {\bf 81}, 064439 (2010).
\bibitem{vondelft}W. Li, A. Weichselbaum, and J. von Delft, arxiv: 1306.5671. 
\bibitem{dechiara} L. Lepori, G. De Chiara, A. Sanpera,  arXiv:1302.5285; G. De Chiara, L. Lepori, M. Lewenstein, A. Sanpera, arXiv:1104.1331.
\bibitem{sorensen} E. Polizzi, F. Mila, and E. S. Sorensen, Phys. Rev. B {\bf 58}, 2407 (1998).
\bibitem{exp_Reny}  A. J. Daley, H. Picher, J. Schachenmayer, and P. Zoller,  Phys. Rev. Lett. {\bf 109}, 020505 (2012);   D. Abanin, and E. Demler, Phys. Rev. Lett. {\bf 109}, 020504 (2012).
\bibitem{note} We note that  at very large $\alpha$ the small eigenvalues produce a log divergence in $S_\alpha$. The correct behavior can be achieved by analyzing $x_1$ (see appendix \ref{large_alpha}).
%\bibitem{supplementary} In the supplementary material we show the edge state, correlation length and string order paramaters in the Cluster-Ising model and the $\lambda-D$ %model, which could determine the topological regions for our numerical results. And we also discussed the R\'{e}nyi entropy in large $\alpha$.
\end{thebibliography}
\end{document}